\documentclass[aps, pre, floatfix]{revtex4-2}

\usepackage{graphicx}
\usepackage{dcolumn}
\usepackage{bm}
\usepackage{physics}
\usepackage{booktabs}
\usepackage{multirow}
\usepackage{braket}
\usepackage{subfigure}
\usepackage{color,hyperref}
\definecolor{darkblue}{rgb}{0.0, 0.0, 1.0}
\hypersetup{colorlinks, breaklinks,
            linkcolor = blue, urlcolor = blue,
            anchorcolor = blue, citecolor = red}

\begin{document}

\title{Steady-State Statistics of Classical Nonlinear Dynamical Systems from Noisy Intermediate-Scale Quantum Devices}

\author{Yash M. Lokare}%
 \email{yash\_lokare@brown.edu}
\affiliation{Department of Physics, Brown University, Providence, Rhode Island 02912, USA}%

\author{Dingding Wei}
\email{ding\_ding\_wei@alumni.brown.edu}
\affiliation{Department of Physics, Brown University, Providence, Rhode Island 02912, USA}


\author{Lucas Chan}
 \email{lucas\_chan@brown.edu}
\affiliation{Department of Physics, Brown University, Providence, Rhode Island 02912, USA}%

\author{Brenda M. Rubenstein}
\email{brenda\_rubenstein@brown.edu}
\affiliation{Department of Chemistry, Brown University, Providence, Rhode Island 02912, USA}%
\affiliation{Department of Physics, Brown University, Providence, Rhode Island 02912, USA}%

\author{J. B. Marston}
\email{Author to whom correspondence should be addressed: marston@brown.edu}
\affiliation{Brown Theoretical Physics Center and Department of Physics, Brown University, Providence, Rhode Island 02912, USA}%

\date{\today}

\begin{abstract}
Classical nonlinear dynamical systems are often characterized by their steady-state probability distribution functions (PDFs). Typically, PDFs are accumulated from numerical simulations that involve solving the underlying dynamical equations of motion using integration techniques. An alternative procedure, direct statistical simulation (DSS), solves for the statistics directly. One approach to DSS is the Fokker-Planck Equation (FPE), which can be used to find the PDF of classical dynamical systems.  Here, we investigate the utility of Noisy Intermediate-Scale Quantum (NISQ) computers to find steady-state solutions to the FPE. We employ the Quantum Phase Estimation (QPE) and the Variational Quantum Eigensolver (VQE) algorithms to find the zero-mode of the FPE for one-dimensional Ornstein-Uhlenbeck problems enabling comparison with exact solutions. The quantum computed steady-state probability distribution functions (PDFs) are demonstrated to be in reasonable agreement with the classically computed PDFs. We conclude with a discussion of potential extensions to higher-dimensional dynamical systems.
\end{abstract}

\maketitle

\section{Introduction}
\label{sec:introduction}

Classical dynamical systems are often characterized by the statistics of a non-equilibrium steady state. These statistics may be accumulated by advancing the system's dynamical equations forward in time using Direct Numerical Simulation (DNS). Such simulations may require long integration times to achieve a desired accuracy and sample rare events. An alternative approach of Direct Statistical Simulation (DSS) seeks solutions of closed-form equations that are derived from the dynamical equations for the statistics themselves.  Because the statistics are often smoother in space and time than the underlying dynamics, DSS is potentially more efficient computationally.  The Fokker-Planck equation (FPE) that describes the time-evolution of the probability distribution function is a well-known example of DSS. Fortunately, the FPE, like quantum mechanics, is linear, enabling the tools of linear algebra to be applied.   In Ref.~\cite{Marston2016}, we have demonstrated that the eigenmode of the linear FPE operator with zero eigenvalue (the ``zero mode'') of the chaotic Lorenz attractor can be found directly (albeit inefficiently) by a classical linear algebra algorithm yielding the probability distribution function  of the attractor. Although an exact solution to the linear FPE can be derived in certain cases, solving the FPE analytically in a general setting is impossible. Under these circumstances, numerical solution of the FPE may still be possible \cite{Hirvijoki2015, Schmidt2009, Caluya2021, Zorzano1999} but must contend with the curse of high dimensionality. This problem potentially could be overcome with the use of a quantum computer \cite{Tennie2024}.    

The central question we wish to address is whether or not there can be an advantage to using Noisy Intermediate-Scale Quantum (NISQ) hardware to solve the FPE. As a first step towards answering this question, in this paper, we demonstrate that quantum computers can accurately find the probability distribution function of a prototypical classical dynamical system.  The illustrative problem we consider here is the Ornstein-Uhlenbeck problem, which describes a one-dimensional  trajectory evolving under the influence of a spatially-dependent velocity $u(x)$ and subjected to random forcing. The system is described by the Langevin equation,
\begin{equation}
    \label{langevin}
    \frac{d\vec{x}}{dt} = \vec{u}(\vec{x}) + \vec{\eta}(t), 
\end{equation}
with Gaussian white noise stochastic forcing $\vec{\eta}(t)$ that parameterizes physics missing from the deterministic velocity $\vec{u}(\vec{x})$ such as the effects of a thermal bath of light particles interacting with a heavy particle \cite{Stinis2020, Hodyss2013}. We study the specific one-dimensional case of $u(x) = - a x - b x^3$ below. 

The FPE respects the fact that dynamical trajectories in an ensemble never terminate and thus total probability is conserved.  Written as the divergence of a flux, it is a linear partial differential equation:
\begin{equation}
    \label{langevin FPE}
    \frac{\partial P(\vec{x}, t)}{\partial t} = -\hat{L}_{\text{FPE}} ~P(\vec{x}, t), 
\end{equation}
where the linear differential Fokker-Planck operator $\hat{L}_{\text{FPE}}$ is given in flux form by:
\begin{equation}
\label{FPE operator}
    \hat{L}_{\text{FPE}}~ P =  \nabla_x \cdot \bigg{(} \vec{u}~ P - \Gamma \vec{\nabla}_x P \bigg{)}\,
\end{equation}
making it clear that total probability is conserved.
The FPE operator is the sum of a drift term $U$ and a diffusion term  with diffusivity given by $\Gamma$, the covariance of the stochastic forcing. The drift term drives a flux of probability density deterministically. The diffusion term captures the stochastic forcing of the dynamical system. It acts to smooth and spread the probability density. Typically, the drift and diffusion terms compete with one another. Note that not only probability but computational difficulty is conserved:  The original nonlinear ordinary differential equation of the Langevin equation is converted to a linear partial differential equation, the FPE, which lives in a space of infinitely higher dimension.  

The steady-state statistics are obtained from the zeromode (or the null space) of the Fokker-Planck operator $\hat{L}_{\text{FPE}}$. The Fokker-Planck operator is semipositive definite, and the null eigenvector of $\hat{L}_{\text{FPE}}$ corresponds to its ground state. Several techniques in the existing literature have been explored extensively to numerically obtain solutions to the FPE including a finite-element method \cite{Wedig2000}, path-integrals \cite{Hegstad1994}, and the discretization of the linear FPE operator on a lattice \cite{Narayanan2006,Marston2016}. 

Other efforts have been made to use quantum computers to simulate the dynamics of linear or weakly nonlinear equations of motion via a set of linear equations \cite{Arrazola.2019,Babbush2023}, but these methods are restricted to regimes of mild non-linearity. Some other groups have also begun to investigate quantum computing the climate \cite{Palmer2023}, or fluid dynamics relevant to components of the climate system \cite{pfeffer2022,pfeffer2023,ingelmann2023}.  An alternative approach to dynamics based upon the linear Koopman -- von Neumann equations is also being investigated \cite{Joseph2020,Slawinska2022,Giannakis2019,Freeman2023} with the statistics accumulated by time integration rather than targeted directly. Quantum algorithms to solve nonlinear stochastic differential equations have also been previously proposed. For instance, Ref. \cite{Surana2023} proposes a method to solve nonlinear stochastic differential equations  using a linear algorithm \cite{Harrow2008}. 
A pressing and thus far unsolved problem is to determine whether or not a quantum advantage exists.  A quantum advantage has been demonstrated for a simple oracle problem \cite{Lidar2023} but an unambiguous demonstration of quantum advantage in the context of a physics simulation problem is still lacking. Other metrics such as energy-to-solution should also be quantified.

Instead of focusing on dynamics, here we investigate the utility of NISQ devices to directly find the statistical steady-state solution of the Fokker-Planck operator as an alternative to traditional classical algorithms. In particular, we employ the Quantum Phase Estimation (QPE) and the Variational Quantum Eigensolver (VQE) algorithms to obtain the steady-state solutions to the one-dimensional FPE. Results for the steady-state probability distribution functions obtained by using these quantum algorithms to the ones obtained via classical numerical diagonalization of the FPE matrix are compared, thereby demonstrating that it is possible to extract the steady-state statistics of classical dynamical systems on near-term quantum hardware. The FPE approach works even for highly nonlinear systems because it turns nonlinear dynamics into a higher-dimensional but linear problem \cite{Venturi.2018,Tennie2024}.  We note that the VQE has been tailored toward the solution of the Fokker–Planck–Smoluchowski eigenvalue problem on near-term quantum hardware for the purpose of estimating the conformational transition rate in a linear chain of rotors with nearest-neighbor interactions \cite{Pravatto2021}. 

The rest of this paper is organized as follows.
Sec. \ref{sec:Model} describes the prototypical nonlinear Ornstein-Uhlenbeck model we use to test our approach, and the analytical solution to the FPE.  The construction of the linear FPE matrix in a reduced dimensional space is presented in Sec. \ref{sec:FPE}. In Sec. \ref{sec:algorithms}, we present the quantum algorithms employed to simulate and/or solve the 1D FPE on near-term quantum hardware.  Sec. \ref{sec:results} reports the results of experiments with IBM NISQ hardware. We discuss the results and related work in Sec. \ref{sec:discussion} and reach some conclusions in Sec. \ref{sec:conclusions}.  

\section{Model Problem: Nonlinear Ornstein - Uhlenbeck Dynamics}
\label{sec:Model}

To investigate DSS using quantum computers, we study a concrete nonlinear Ornstein-Uhlenbeck problem for a particle moving in just one spatial dimension with non-linear velocity given by
\begin{eqnarray}
u(x) = - a x - b x^3\
\label{u(x)}
\end{eqnarray} 
with $b \geq 0$ for stability.
The dynamics are specified by
\begin{equation}
    \label{1D ornstein uhlenbeck}
    \frac{dx}{dt} = u(x) + \eta(t), 
\end{equation}
where $\eta(t)$ is a Gaussian white noise term that satisfies
\begin{equation}
    \label{white noise}
    \langle \eta(t)~ \eta(t') \rangle = 2 \Gamma~ \delta(t - t'), 
\end{equation}
with covariance $\Gamma$.
For $a < 0$ the model is bistable with the velocity vanishing at two locations $x = \pm \sqrt{-b/a}$.  Subjected to stochastic forcing, the system makes transitions between these two metastable states.  The problem we consider is thus a special case of Kramer's escape problem \cite{Kramer1940}.  It can be used as a highly simplified model for problems as diverse as isolated crystalline vacancies and transitions between two distinct climate states. For $a > 0$ there is only one stable state and the model captures some effects of nonlinearity. The model is of sufficiently low dimensionality that it can be studied with current-day NISQ computers.  As the hardware and algorithms become more capable, more realistic models will be accessible using generalizations of the methods presented here (we comment on future prospects in Section \ref{sec:conclusions}).

The corresponding FPE for the stationary (invariant) measure $P(x)$ is:
\begin{equation}
    \label{1D OU FPE}
    (ax + bx^3) P(x) + \Gamma \frac{d}{dx} P(x) = 0\ . 
\end{equation}
Note that the equation, and thus $P(x)$, is invariant under rescaling $a$, $b$, and $\Gamma$ by a common factor as this only changes the rate of approach to the steady state.  
It is easily verified that the solution is given by:
\begin{equation}
    \label{exact solution}
    P(x) = N ~\exp\biggl(-\frac{a x^2}{2 \Gamma} - \frac{b x^4}{4 \Gamma}\biggr), 
\end{equation}
where $N$ is the normalization factor determined by imposing the $L_1$ norm:
\begin{eqnarray}
\int_{-\infty}^\infty P(x)~ dx = 1.\
\label{normalization}
\end{eqnarray}
(Note that it is the wavefunction itself that is normalized, rather than its square as in quantum mechanics.)  
Low-order moments such as $\langle x^2 \rangle$ may be easily computed from Eq. \ref{exact solution} by numerical integration. 

\section{The FPE operator in a finite basis of Hermite polynomials}
\label{sec:FPE}

To employ a quantum computer, the infinite-dimensional linear Fokker-Planck operator $\hat{L}_{\text{FPE}}$ must be projected into a finite basis of dimension $N$ that is a good approximation to the infinite-dimensional space. Finite-difference discretization is possible but inefficient \cite{Chakraborty2015,Marston2016}. Instead, here we choose to work in a basis of quantum harmonic oscillator energy eigenstates with position space amplitudes given by:
\begin{equation}
    \label{hermite}
    \langle x | n \rangle = \frac{\sqrt{\ell}}{\pi^{1/4} \sqrt{2^n n!}} ~H_n(x/\ell) ~e^{-x^2/2 \ell^2}\ . 
\end{equation}
Here, $H_n(x/\ell)$ is the $n$th-order Hermite polynomial and $\ell$ is a characteristic length scale that can be tuned to optimize the basis. (For the case $b = 0$, the exact PDF is a Gaussian described by the $n = 0$ basis state with $\ell = \sqrt{\Gamma / a}$.) The states form an orthonormal basis with inner product $\braket{m | n} = \delta_{m n}$. The matrix elements of $\hat{L}_{\text{FPE}}$ in this basis are simply $L_{m n} = \bra{m} \hat{L}_{\text{FPE}} \ket{n}$. Note, however, that when $b > 0$ in Eq. \ref{u(x)}, the PDF decays at large $|x|$ as $\exp[- b x^4 / \Gamma]$, more rapidly than the basis states of Eq. \ref{hermite} (see Eq. \ref{exact solution}; alternatively the WKB approximation also shows this to be the case).  Therefore, any finite truncation in this basis must break down at sufficiently large values of $|x|$.  

To calculate the matrix elements, we introduce lowering and raising operators $\hat{a}$ and $\hat{a}^{\dagger}$ : 
\begin{eqnarray}
\label{ladder operators}
        \hat{a} \equiv \frac{1}{\sqrt{2}} ~\biggl(\frac{x}{\ell} + \ell ~\frac{d}{d x}\biggr), \\
        \hat{a}^{\dagger} \equiv \frac{1}{\sqrt{2}} ~\biggl(\frac{x}{\ell} - \ell ~\frac{d}{d x}\biggr), 
\end{eqnarray}
that have well-known actions upon the basis states:
\begin{eqnarray}
    \label{ladder properties}
        \hat{a} \ket{n} &=& \sqrt{n} \ket{n - 1}, \\
        \hat{a}^{\dagger} \ket{n} &=& \sqrt{n + 1} \ket{n + 1}, 
\end{eqnarray}
Then, $x$ and $d/d x$ are given by:
\begin{eqnarray}
    \label{position momentum operators}
        x \equiv \frac{\ell}{\sqrt{2}} ~(\hat{a} + \hat{a}^{\dagger}), \\
        \frac{d}{d x} \equiv \frac{1}{\sqrt{2} \ell}~(\hat{a} - \hat{a}^{\dagger}). 
\end{eqnarray}

The zero-mode of the linear Fokker-Planck operator,
\begin{equation}
    \label{FPE steady state}
    \hat{L}_{\text{FPE}} ~P(x) = 0,
\end{equation}
is the steady-state PDF we seek, however $\hat{L}_{\text{FPE}}$ is neither of finite dimension nor generally self-adjoint.  To use a quantum computer to find the zero-mode, the following matrix ``Hamiltonian'' $H$ with matrix elements
\begin{equation}
    H_{m n} = \sum_{p = 0, 2, 4, \ldots}^{N~ {\rm states}} L_{p m}^* L_{p n}\ .
\end{equation}
is introduced. The sum is over only $N$ even basis states, building in the symmetry of the PDF under $x \rightarrow -x$.  $H_{m n}$ has the same zero-mode as $L_{m n}$ but has the virtue of being Hermitian \cite{Marston2016}.  (Alternatively, one could form $\hat{H} = \hat{L}^\dagger_{\text{FPE}} \hat{L}_{\text{FPE}}$ and then project the differential operator into the finite basis; the result is not exactly identical to $H_{m n}$ due to the fact that $\hat{L}_{\text{FPE}}~ \ket{2N-2}$ generates state $\ket{2N}$ that lies outside of the subspace.) The spectrum of $H$, like $\hat{L}_\text{FPE}$, is non-negative.

The ground state $\ket{\psi_0}$ of $H$ may be written in terms of amplitudes $b_n$ for the basis states
\begin{equation}
\ket{\psi_0} = \sum_{n = 0, 2, 4, \ldots}^{N~ {\rm states}} b_n \ket{n}\
\end{equation}
The probability density is then given approximately by
\begin{equation}
P(x) \approx \langle x | \psi_0 \rangle = \sum_{n = 0, 2, 4, \ldots}^{N~ {\rm states}} b_n  \frac{\sqrt{\ell}}{\pi^{1/4} \sqrt{2^n n!}} ~H_n(x/\ell) ~e^{-x^2/2 \ell^2}\ .
\label{expansion}
\end{equation}
Eigenstates are defined only up to an overall complex-valued constant (they are ``rays'' in a Hilbert space.)  This constant must be chosen to both ensure that the zero-mode may be interpreted as a normalized probability that is real-valued, non-negative ($P(x) \geq 0$), and normalized per Eq. \ref{normalization}. Positivity is violated for any finite basis truncation $N < \infty$ because the behavior at $|x| \rightarrow \infty$ is dominated by the highest Hermite polynomial in the truncation which has $2N-2$ nodes, but the violations are suppressed by the $\exp[-x^2 / 2 \ell^2]$ prefactor and thus will be small in magnitude. 

The ground state amplitudes $b_n$ may be found either by classical numerical diagonalization or by the quantum methods as described in the next section.
Contrary to intuition, a moment such as $\langle x^2 \rangle$ is \emph{not} given by the quantum expectation value  $\langle \psi_0 | x^2 | \psi_0 \rangle$ but rather by
\begin{equation}
\langle x^2 \rangle = \int_{-\infty}^\infty x^2 P(x) dx~ \approx \sum_{n = 0, 2, 4, \ldots}^{N~ {\rm states}} b_n \int_{-\infty}^\infty x^2 \frac{\sqrt{\ell}}{\pi^{1/4} \sqrt{2^n n!}} ~H_n(x/\ell) ~e^{-x^2/2 \ell^2} dx\ 
\label{expectation}
\end{equation}
assuming that the $b_n$ amplitudes have been normalized such that $\langle 1 \rangle = 1$.  The integrals on the RHS of Eq. \ref{expectation} may be evaluated analytically, giving a direct relationship between the moment and the amplitudes $b_n$.  

\begin{figure}[ht]
\includegraphics[width = 0.8 \columnwidth]{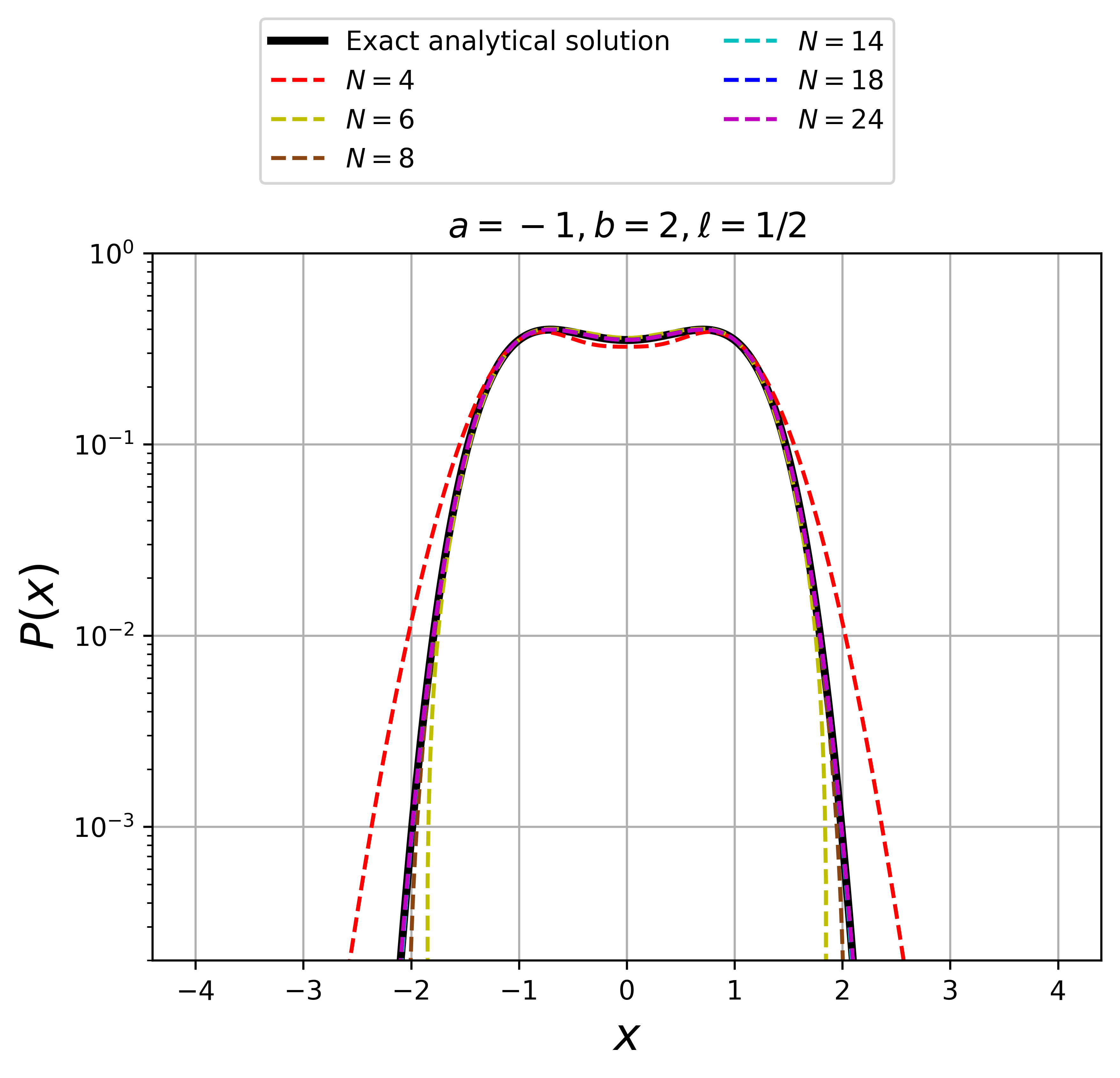}
\caption{\label{finite truncation plots} Steady-state probability distribution functions obtained via the exact analytical solution and classical numerical diagonalization for differently-sized truncations of the basis.  Results for $N = 14$, $18$ and $24$ lie underneath the exact analytical solution in the displayed range.}
\end{figure}

To test the viability of the reduced basis, we find the steady-state probability distribution functions for different Hermite polynomial basis sizes via numerically exact classical diagonalization and compare the PDFs against the exact analytical solution for the following model FPE parameters: $a = -1, ~b = 2, ~\Gamma = 1, ~\ell = 1/2$; see Fig. \ref{finite truncation plots}. For large basis sizes, the steady-state probability distribution functions obtained via classical diagonalization agree well with the exact analytical solution over the studied range.  For smaller truncations, discrepancies are visible in the tails of the distributions due to the aforementioned fact that the basis functions do not decay fast enough. 

\section{Quantum algorithms}
\label{sec:algorithms} 
Here, we briefly describe the Quantum Phase Estimation (QPE) and Variational Quantum Eigensolver (VQE) algorithms used for the computational experiments in Sec. \ref{sec:results} below.

\subsection{QPE}
\label{sec:QPE}

QPE estimates the eigenvalues of an unitary matrix $U$ with matrix elements given by $U_{mn} = (\exp[i H])_{mn}$ \cite{Kitaev1995}. The eigenvalue associated with eigenvector $\ket{u}$ is given by $e^{2\pi i \phi}$ for some $\phi \in [0, 2 \pi]$. The QPE algorithm estimates the value of the phase $\phi$ as $\phi = 2 \pi 0.j_1j_2 \ldots j_n$ in binary representation, with $n$ denoting the degree of precision. 
The QPE circuit is composed of two registers -- the precision qubit register and the query qubit register. The precision qubits are used to represent the eigenvalues of $\hat{U}$ in binary format, whereas the query qubits store information about the corresponding eigenvectors. 
We run QPE using Qiskit \cite{Qiskit} on both the local simulator and the IBMQ Montreal machine, a 27-qubit quantum processing unit (QPU). The number of query qubits is $n = \log_2(N)$, where $N$ denotes the dimension of $\hat{H}$. $7$ qubits are used for the precision qubit register and the $N = 8$ basis size requires $3$ query register qubits. The optimization level of the QPE circuit is set to $1$. 

\subsection{VQE}
\label{VQE}

The VQE algorithm is a hybrid classical-quantum algorithm used to find the ground state energy of a quantum system \cite{Peruzzo2014, Wang2018}. The VQE leverages the variational principle, which states that the expected value of the Hamiltonian is always greater than or equal to the ground state energy. The algorithm iteratively refines the trial wavefunction to minimize this expected value, ideally approaching the true ground state energy as the number of variational parameters is increased. 

Expressed mathematically, the cost function  is the expectation value of the Hamiltonian:
\begin{equation}
    \label{VQE cost}
    C(\vec{\theta}) = \bra{\psi(\vec{\theta})} H \ket{\psi(\vec{\theta})}, 
\end{equation}
where the Hamiltonian $H$ is decomposed into a linear combination of strings of Pauli operators $P_i$ as $H = \sum_{i=1}^{4^n} c_i P_i$.

We implement VQE on the Qiskit platform \cite{Qiskit}. For the VQE experiments, the hardware-efficient {\tt RealAmplitudes} variational ans\"atz is used with $4$ repetition layers. The Implicit Filtering (IMFIL) \cite{Kelley2011} and the Simultaneous Perturbation Stochastic Approximation (SPSA) \cite{Spall1992} optimization algorithms are used to run VQE experiments on the IBMQ Manila machine, a 5-qubit QPU. The VQE experiments are run with a different number of measurement shots and classical optimization iterations across different cases to explore convergence. We use the {\tt Estimator} primitive available in Qiskit \cite{Qiskit} to compute expectation values in VQE and the {\tt Statevector} class to read out the state vector of the variational ans\"atz once the optimization loop terminates. In all VQE experiments, a random initial state is employed, where the parameters for the variational ans\"atz are drawn from an uniform distribution bounded within the interval $[-\pi, \pi]$. The optimization level of the variational ans\"atz is set to the highest possible value of 3 and error mitigation protocols are employed to suppress the effects of hardware noise in the VQE experiments (see below). 

\subsection{Error Mitigation}
\label{error mitigation}

Zero noise extrapolation (ZNE) \cite{Ying2017, Temme2017, Kandala2019, Majumdar2023} is an error mitigation technique in which the expectation value of an observable is estimated at different noise levels. The ideal expectation value is then inferred by extrapolating the measurement results to the zero-noise limit.

Twirled Readout Error Extinction (TREX) is an error correction protocol that suppresses the effect of measurement errors for the estimation of the expectation values of Pauli observables \cite{Bravyi2021, Berg2022}. The scheme improves the accuracy of quantum measurements by applying randomized twirling operations to quantum states before measurement, distributing errors uniformly across different outcomes. TREX mitigates readout errors by analyzing and characterizing these distributed errors through multiple measurements.

\section{NISQ Experiments}
\label{sec:results}

In this section, results for the invariant measure obtained via the implementation of QPE and VQE on near-term quantum hardware are presented. We compare the steady-state PDFs obtained via QPE and VQE to the PDFs obtained via classical numerical diagonalization. The Qiskit software versions used to run experiments are as follows: {\tt qiskit}: 1.1.1, {\tt qiskit-ibm-runtime}: 0.25.0, {\tt qiskit-aer}: 0.12.0, {\tt scikit-quant}: 0.8.2, and {\tt qiskit-algorithms}: 0.3.0. 

\subsection{QPE}
\label{QPE section}

Here, we motivate the use of the VQE algorithm by first demonstrating that QPE by itself is insufficient to recover the desired zeromode. For this purpose, the following model parameters are used: $a = -1, b = 2, \Gamma = 1, \ell = 1/2$. A basis size of $N = 8$ is employed for the experiments, yielding a query qubit register size of three. Seven precision qubits are used to construct the precision qubit register. The amplitudes of the zeromode are obtained by taking the square root of the probabilities of collapsing into each Hermite polynomial basis state.  Thus, QPE does not provide information about the signs of the amplitudes. Signs can be added by hand (knowing the exact classical solution), but of course, there is no need for quantum computation if this information is used.  

\begin{figure}[ht]
\includegraphics[width = 0.8 \columnwidth]{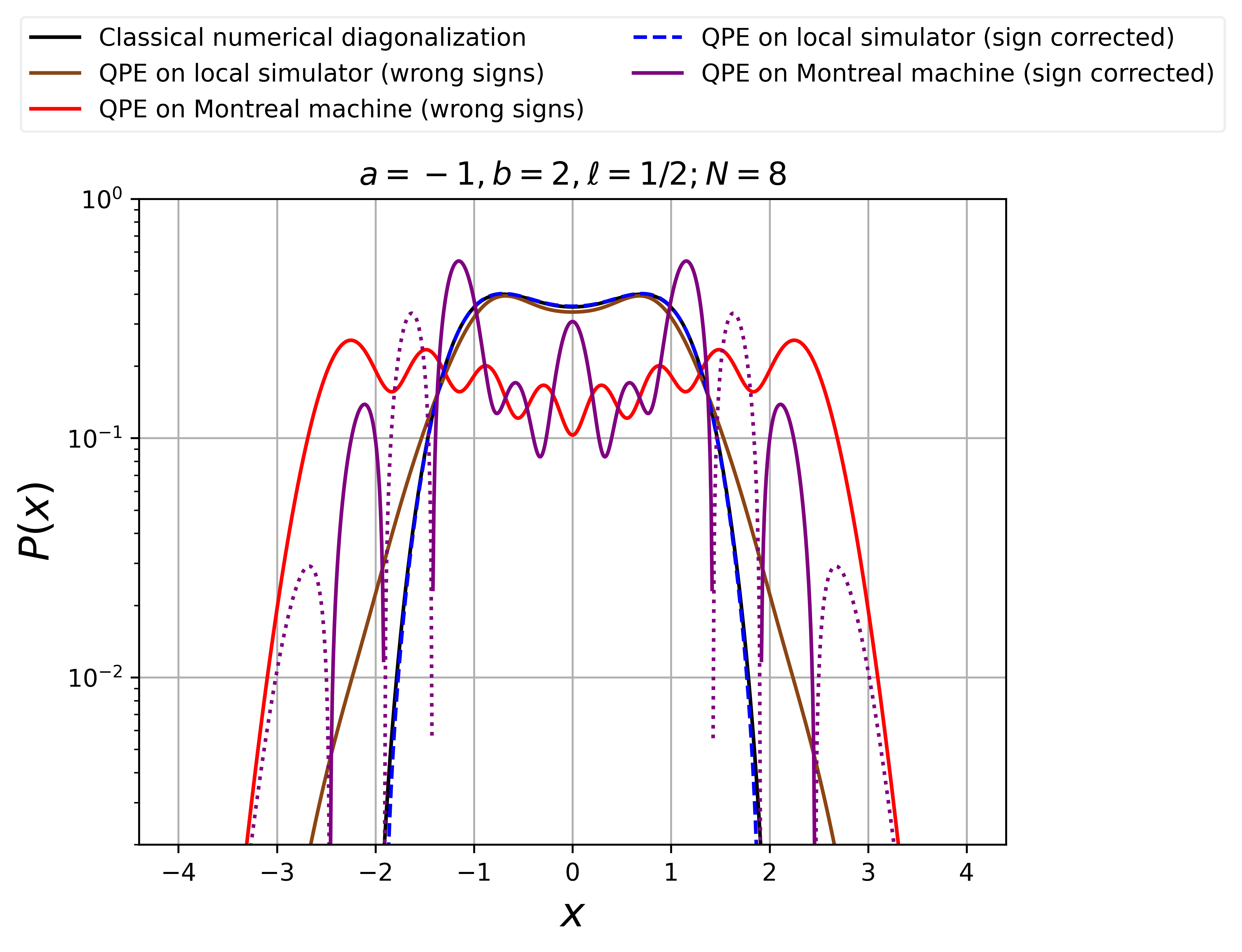}
\caption{\label{QPE} The steady-state PDFs obtained via classical numerical diagonalization and QPE. Classical numerical diagonalization (black line) should be considered the ``truth'' against which the quantum experiments with NISQ hardware are compared.  PDFs obtained from a QPE ground state with all positive amplitudes and with sign-corrected amplitudes are shown. Regions where the probabilities become  negative are indicated with dotted purple lines.}
\end{figure}

The steady-state PDFs obtained via classical numerical diagonalization of the FPE matrix and QPE are shown in Fig. \ref{QPE}.  An accurate approximation to the classically computed PDF with the sign-corrected QPE zeromode recovered on the local simulator is obtained. We note, however, that poor approximations to the classically computed PDF are obtained with the QPE zeromode (with wrong signs on the zeromode amplitudes) recovered on both the local simulator and the IBMQ Montreal machine. The steady-state PDF with the sign-uncorrected zero-mode recovered on the IBMQ Montreal machine is significantly less accurate as hardware noise degrades the performance of QPE.  

\subsection{VQE with Zero Noise Extrapolation}
\label{VQE section}

In this section, results for the steady-state PDFs obtained via VQE with ZNE are presented. Accurate approximations to the classical diagonalization results are obtained when VQE was run on the ideal simulator. Here we focus on experimental VQE results obtained using the IBM Manila machine. The following model FPE parameters are used in the VQE experiments: $a = \pm 1, b = 2, \ell = 1/2$ (with $\Gamma = 1$).

\subsubsection{Results for a = 1}
\label{Positive a}

\begin{figure*}[ht]
    \centering
    \subfigure[]{
        \includegraphics[width=0.48\textwidth]{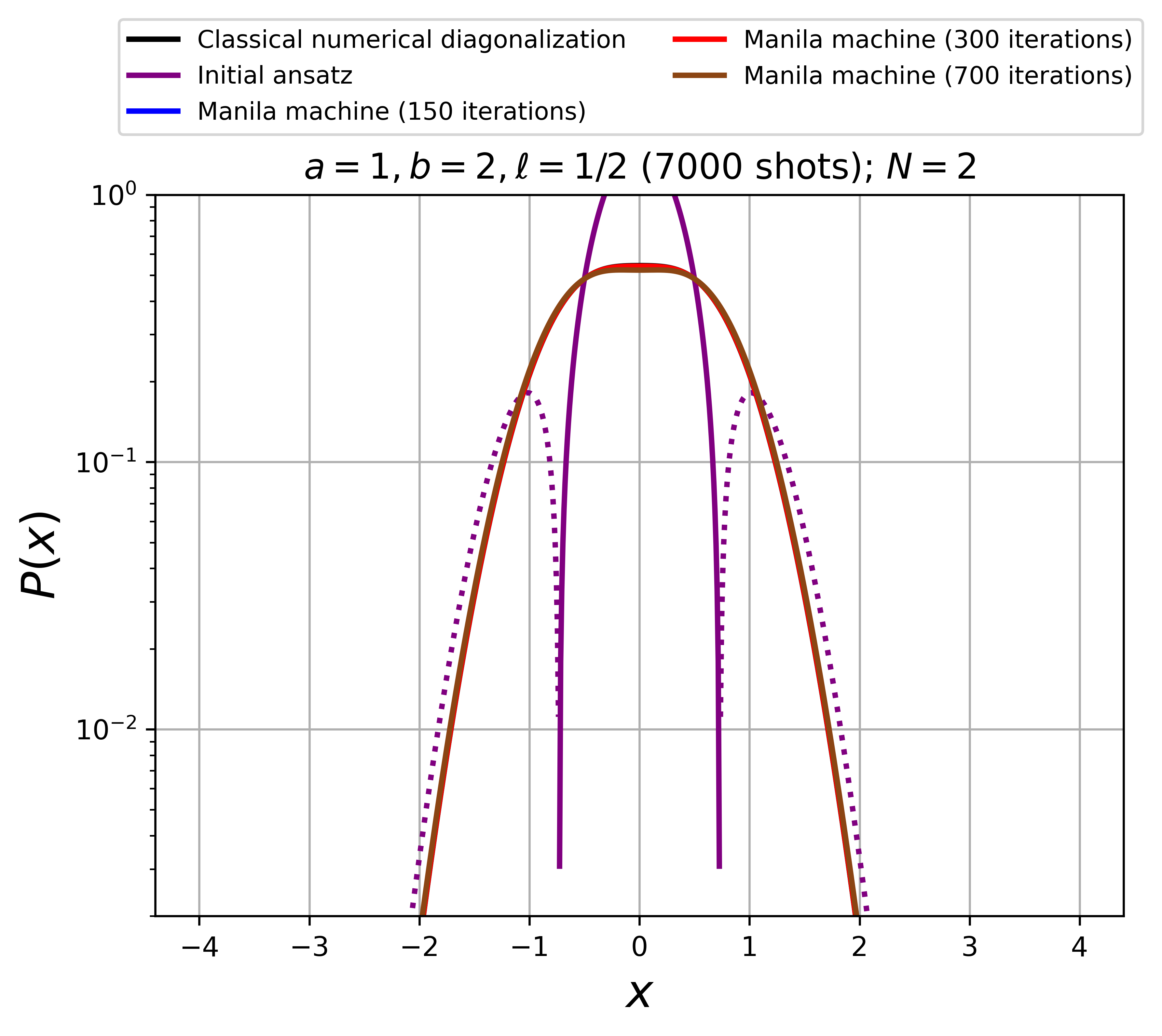}
        \label{$N = 2$ ($a = 1$)}
    }
    \hfill
    \subfigure[]{
        \includegraphics[width=0.48\textwidth]{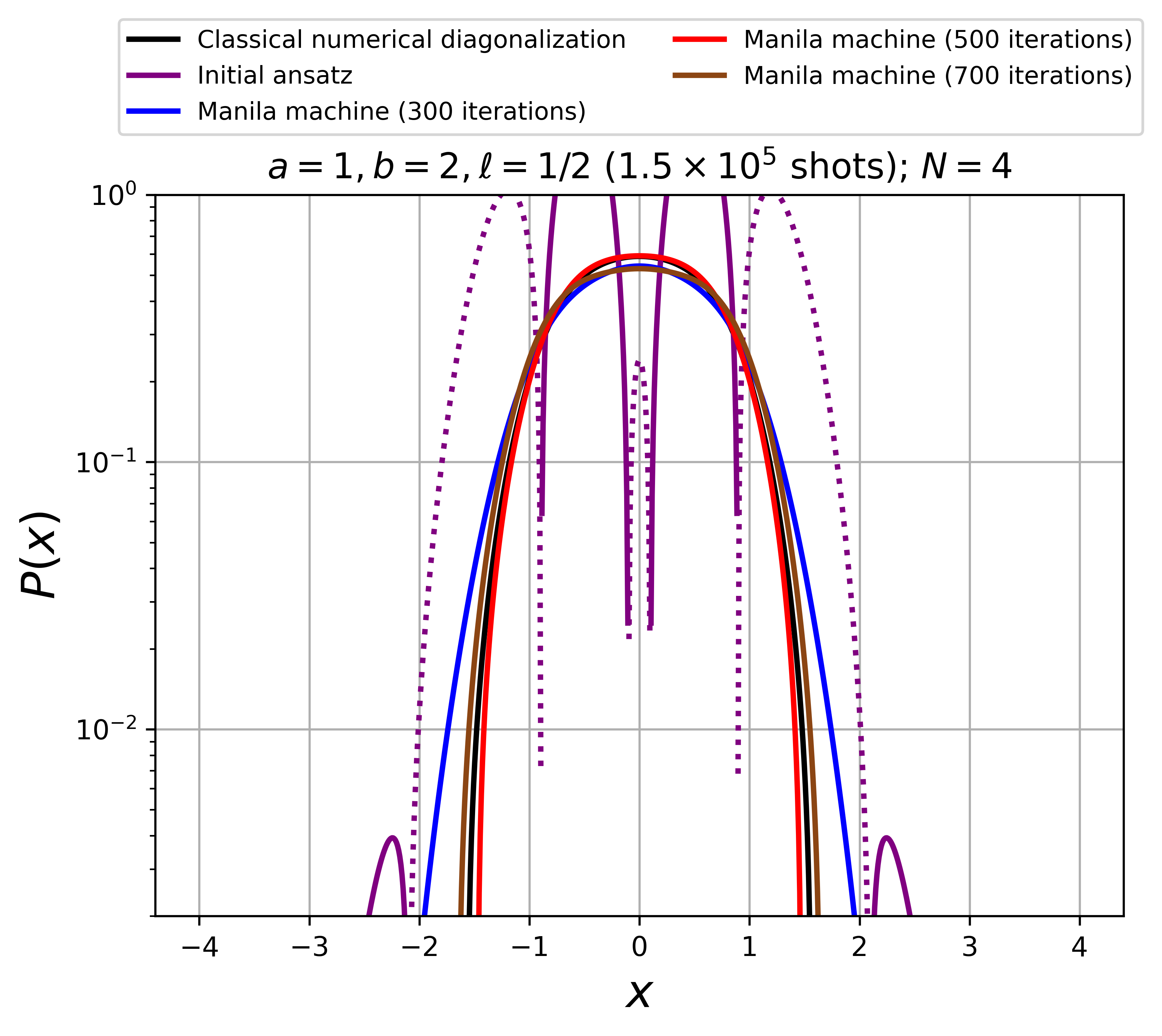}
        \label{$N = 4$ ($a = 1$)}
    }
    \vspace{0.5cm}
    \centering
    \subfigure[]{
        \includegraphics[width=0.48\textwidth]{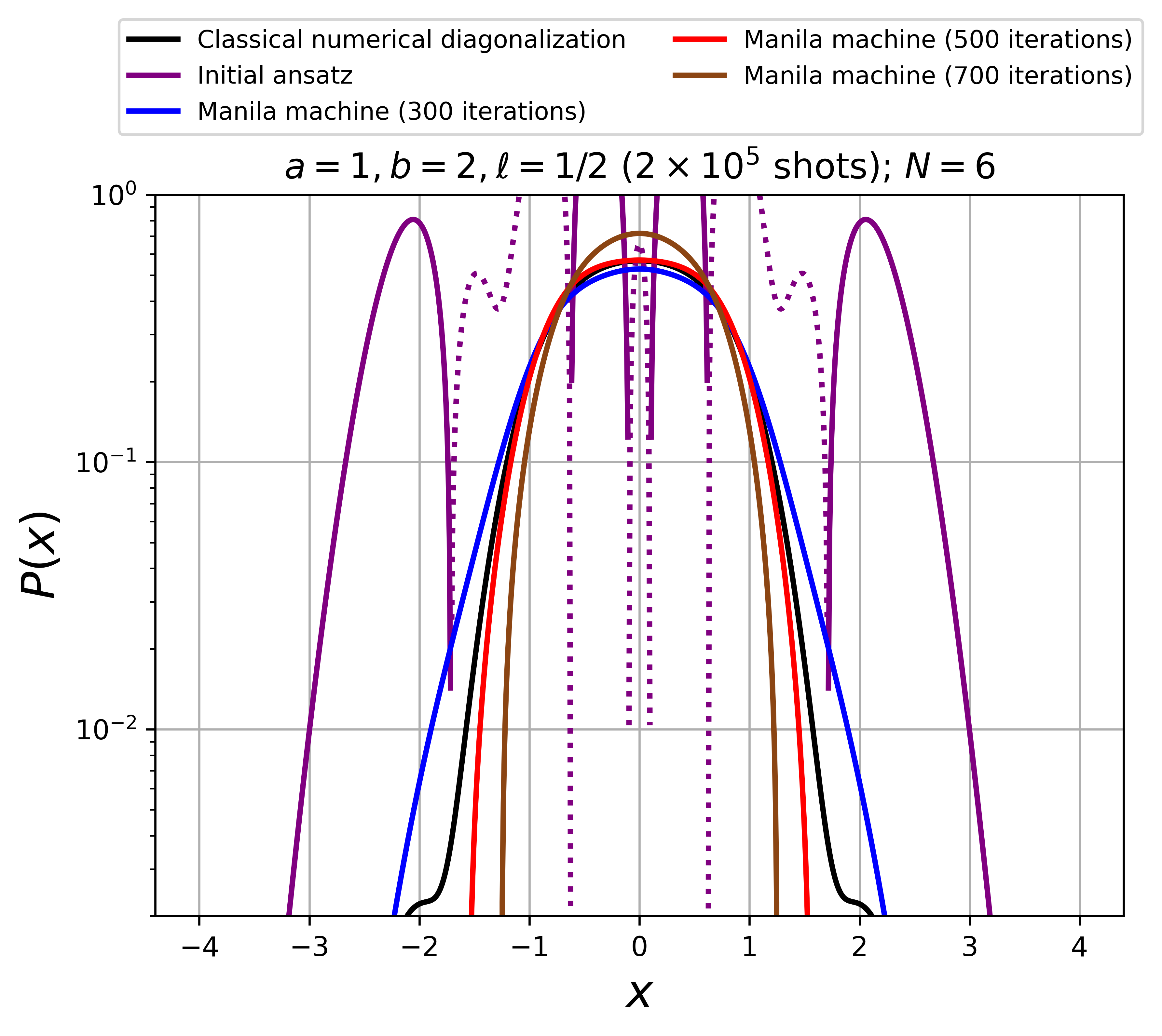}
        \label{$N = 6$ ($a = 1$)}
    }
    \caption{Steady-state PDFs obtained via classical numerical diagonalization and VQE experiments on the IBMQ Manila machine for FPE matrices of different dimension using ZNE for truncations $N = 2$ (a), $4$ (b) and $6$ (c).  Classical numerical diagonalization (black) should be viewed as the ``truth'' against which the quantum experiments are compared.
    Regions where the probabilities go negative are indicated with dotted purple lines. In (a) the black and red lines lie underneath brown.}
    \label{positive a}
\end{figure*}

The steady-state PDFs for the case $a = 1$ obtained via numerically exact classical numerical diagonalization and VQE experiments with ZNE for $N = 2, 4$, and $6$ are shown in Fig. \ref{positive a}. For $N = 2$ and $N = 4$, the IMFIL optimizer is used to perform a classical optimization of the variational parameters. For $N = 6$, however, the SPSA optimizer is used instead as it was observed that the IMFIL optimizer yields erroneous results for the zero-mode, and thus the steady-state PDF. 

Good quantitative agreement between the classical and quantum-computed steady-state PDFs is obtained for the $N = 2$ case for 150, 300, and 500 iterations (relative errors for $\langle x^2 \rangle$ on the order of $0.4-2 \%$ are obtained). The steady-state PDF obtained via VQE with 500 iterations is slightly less accurate as compared to the PDFs obtained with 150 and 300 iterations, 
Similar observations may be made for the $N = 4$ and $N = 6$ cases, where accurate approximations to the classically computed PDF are obtained via VQE with 500 iterations, with relative errors for $\langle x^2 \rangle$ being on the order of 9\% in the worst-case scenario (with 300 iterations, the steady-state solution(s) obtained are significantly less accurate), although with 700 iterations, the accuracy again drops. Going beyond $N = 6$ to $N = 8$, we find that results deteriorate markedly, suggesting that improvements to the variational algorithms and/or hardware are necessary for higher-dimensional problems. 

\begin{figure}[ht]
\includegraphics[width = 0.8 \columnwidth]{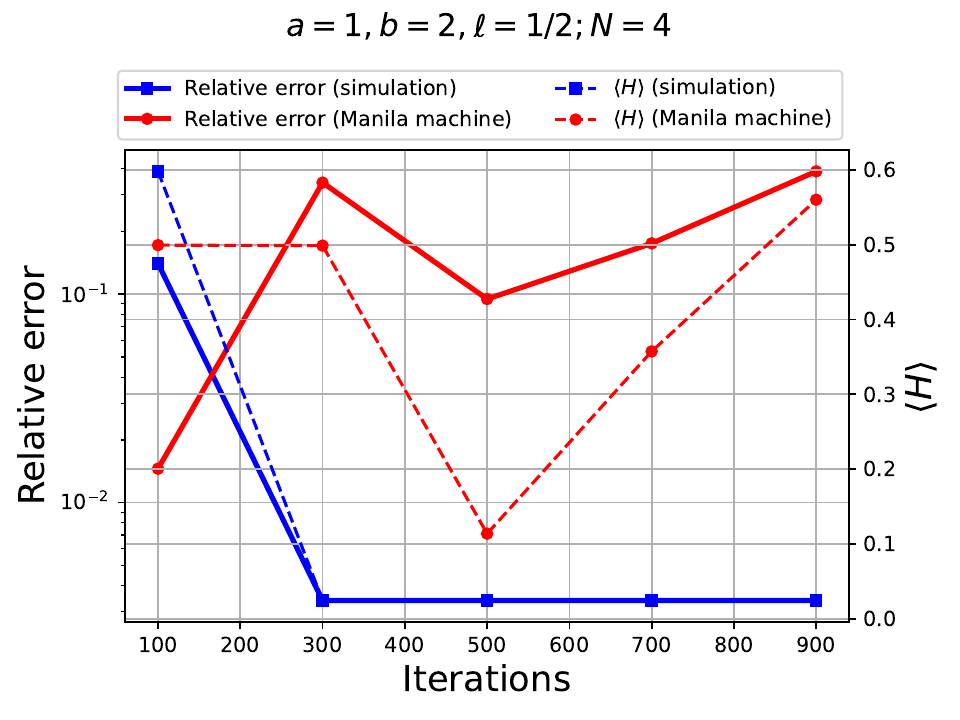}
\caption{\label{Hardware-simulation comparison} Relative error in $\langle x^2 \rangle$ (solid lines) and $\langle H \rangle$ (dashed lines) for $a = 1$ and $N = 4$ as a function of the number of iterations. The IMFIL optimizer is used for both noiseless simulations (blue) and hardware experiments on Manila (red).}
\end{figure}

Some understanding of the performance of VQE may be gleaned from Figure \ref{Hardware-simulation comparison} which displays the relative error in $\langle x^2 \rangle$, defined as $\frac{ |\langle x^2 \rangle_\text{VQE} - \langle x^2 \rangle_\text{exact}| }{ \langle x^2 \rangle_\text{exact} }$, and $\langle H \rangle$ for the $a = 1$ and $N = 4$ case. It is evident that deviations in $\langle H \rangle$ away from the exact value of $0$ correspond to deviations in the accuracy of the PDF.  Furthermore the 
hardware experiments do not show uniform convergence towards the most accurate result but rather fluctuate, suggesting that noise may be responsible. 

\subsubsection{Results for a = -1}
\label{Negative a}

\begin{figure*}[ht]
    \centering
    \subfigure[]{
        \includegraphics[width=0.48\textwidth]{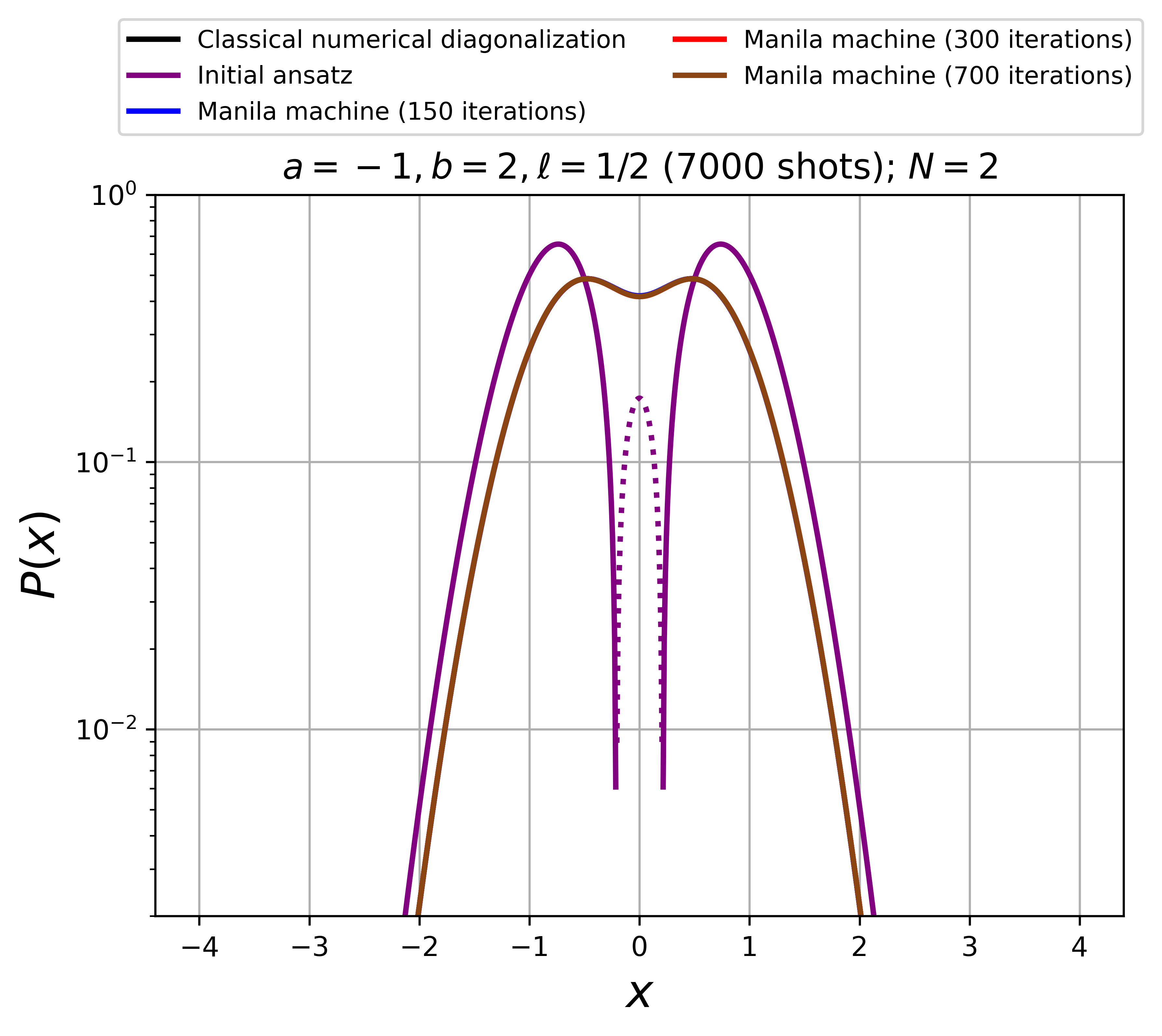}
        \label{$N = 2$ ($a = -1$)}
    }
    \hfill
    \subfigure[]{
        \includegraphics[width=0.48\textwidth]{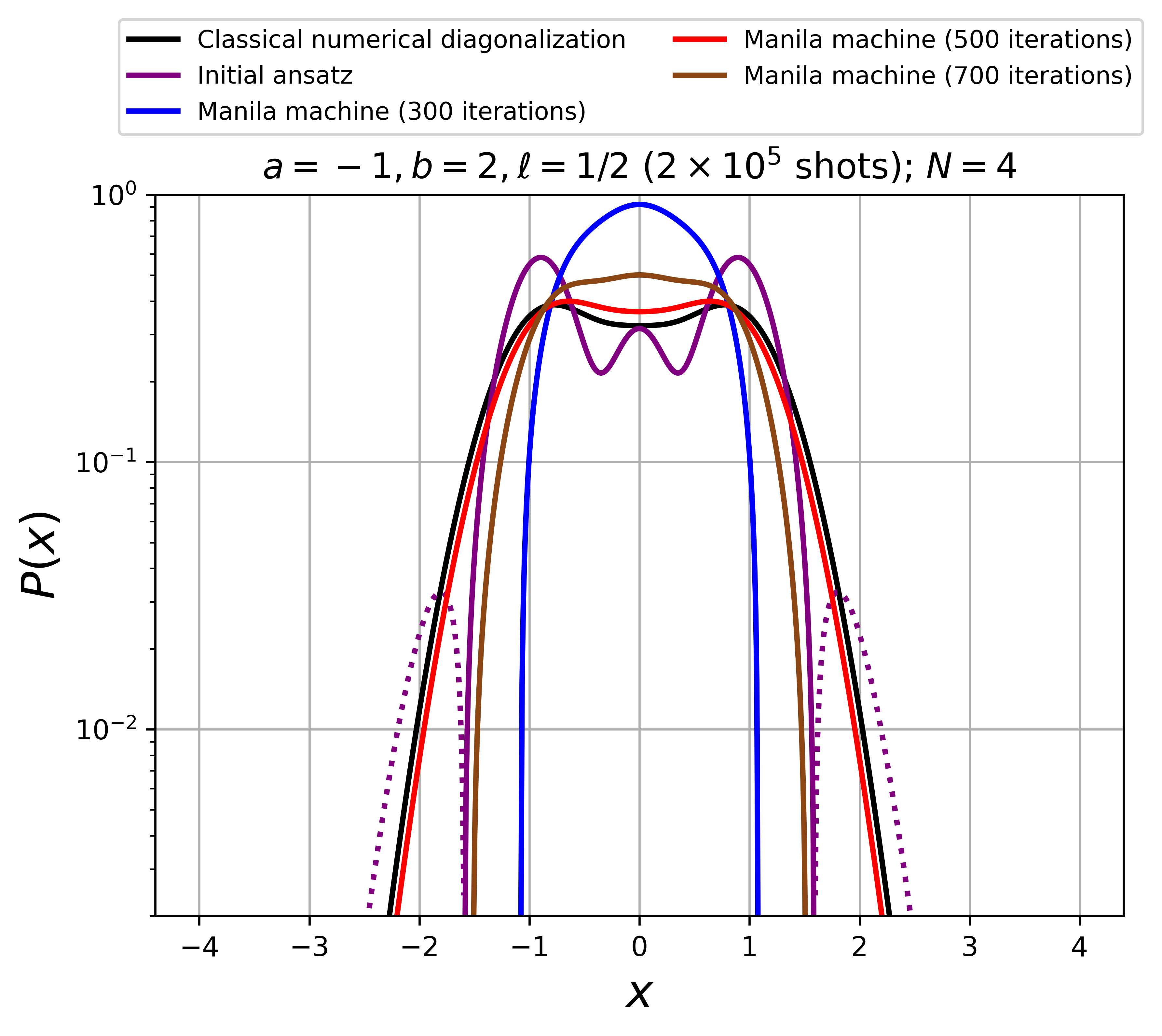}
        \label{$N = 4$ ($a = -1$)}
    }
    \vspace{0.5cm}
    \centering
    \subfigure[]{
        \includegraphics[width=0.48\textwidth]{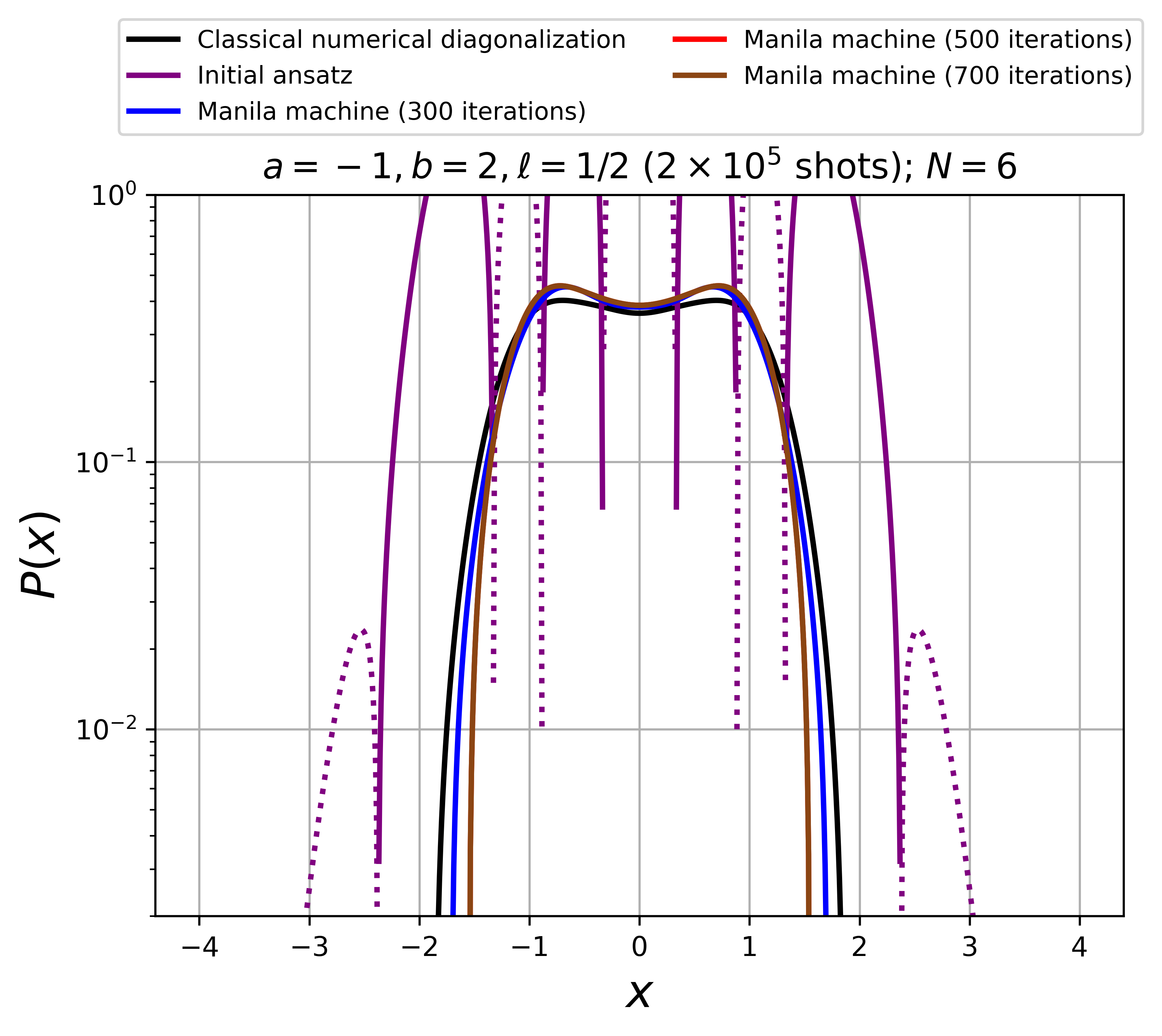}
        \label{$N = 6$ ($a = -1$)}
    }
    \caption{Steady-state PDFs obtained via classical numerical diagonalization and VQE experiments on the IBMQ Manila machine for FPE matrices of different dimension using ZNE for truncations $N = 2$ (a), $4$ (b) and $6$ (c). Regions where the probabilities become negative are indicated with dotted purple lines. In (a) the black, blue and red lines lie underneath brown.}
    \label{negative a}
\end{figure*}

A more interesting case is obtained for $a = -1$ as the PDF has two peaks. This system can be used as a model of transitions between two local equilibria. A paradigmatic example is the stochastic (bistable) Schl\"ogl model \cite{Schlogl1972, kabengele2024modeling}. Steady-state PDFs obtained via classical numerical diagonalization and VQE experiments with ZNE for $N = 2, 4$, and $6$ are shown in Fig. \ref{negative a}. 
Good quantitative agreement is obtained between the classical and quantum-computed steady-state PDFs for the $N = 2$ case (relative errors for $\langle x^2 \rangle$ being on the order of 0.2\%), with little to no visible deviations observed between the PDFs obtained via VQE for optimization routines of different length. 
For $N = 2$ and $N = 4$, significantly less accurate approximations to the classically-computed PDF(s) are obtained via VQE with 300 iterations, although with 500 iterations, better approximations to the steady-state solution(s) are recovered. It is furthermore observed that, with 700 iterations, the accuracy of the steady-state PDF for the $N = 4$ case decreases again, although no visible deviations are observed for the $N = 6$ case. 

\subsection{VQE with Twirled Readout Error Extinction}
\label{VQE with TREX}

To test noise mitigation with TREX, VQE is run for the same number of optimization iterations, the same choice of the variational ans\"atz and classical optimizer, and the same number of measurement shots as ZNE to ensure a fair comparison. 

\begin{figure}[ht]
  \centering
  \subfigure[]{
    \includegraphics[width=0.48\linewidth]{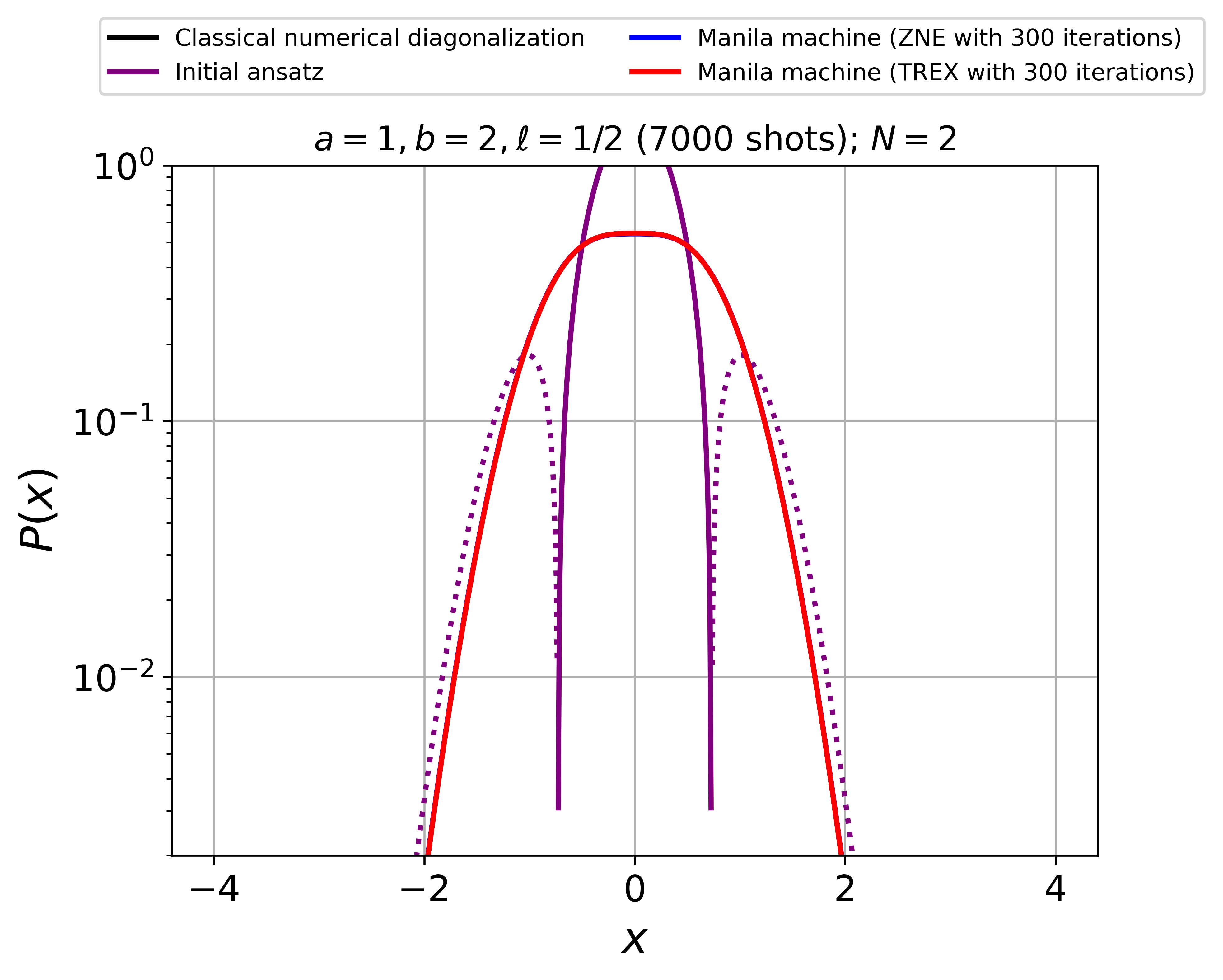}
    \label{n = 2 positive}
  }
  \subfigure[]{
    \includegraphics[width=0.48\linewidth]{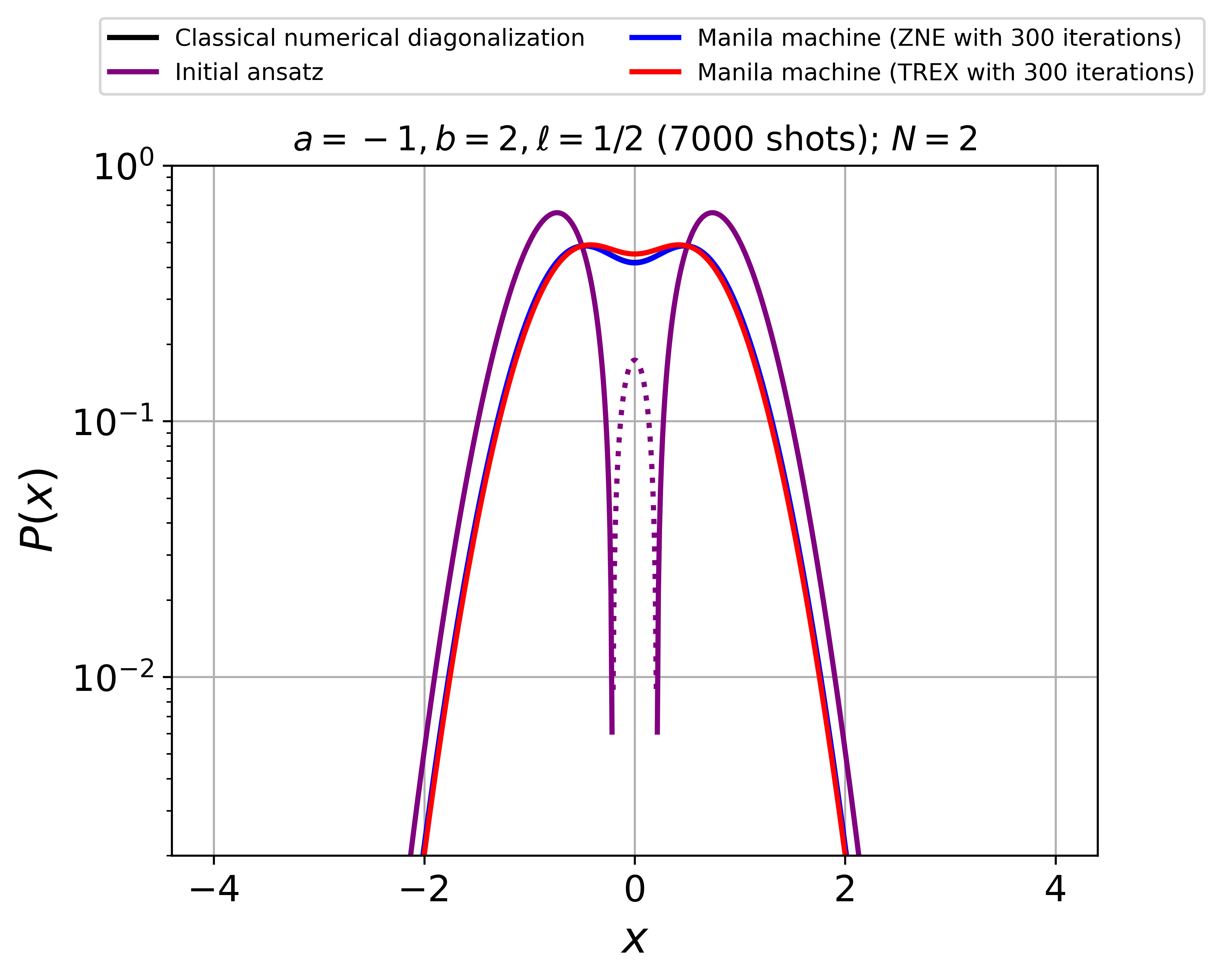}
    \label{n = 2 negative}
  }
  \\
  \subfigure[]{
    \includegraphics[width=0.48\linewidth]{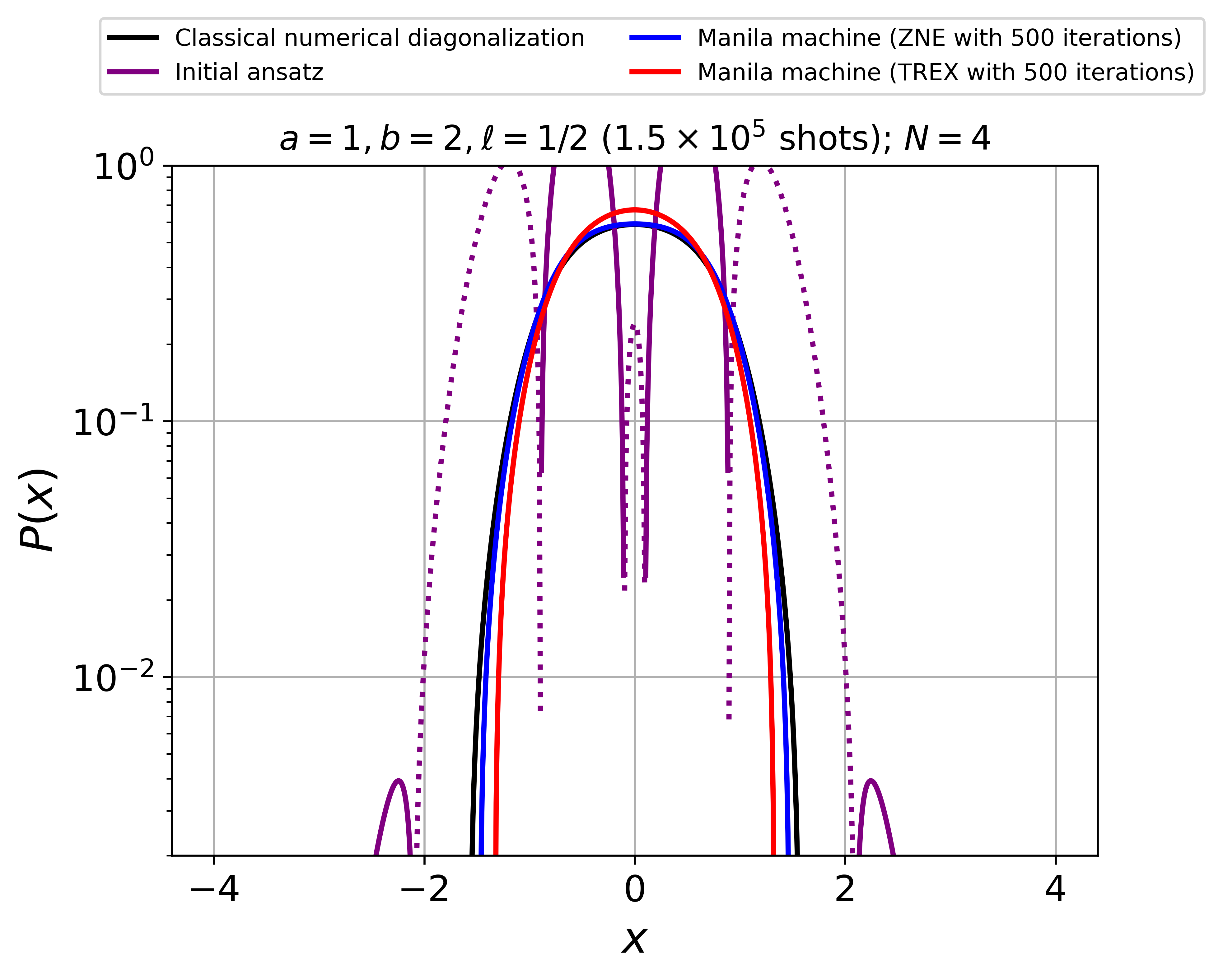}
    \label{n = 4 positive}
  }
  \subfigure[]{
    \includegraphics[width=0.48\linewidth]{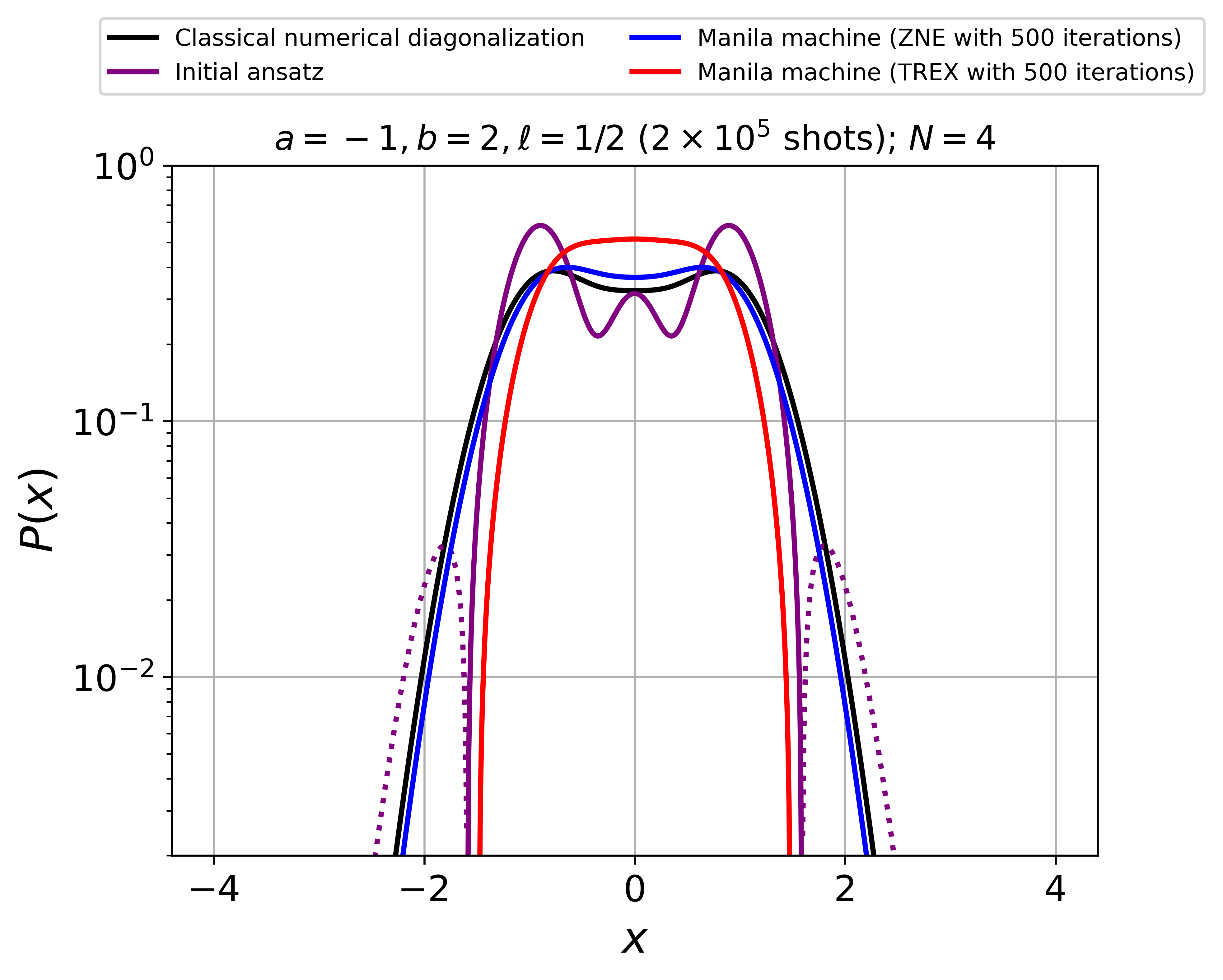}
    \label{n = 4 negative}
  }
  \caption{Steady-state PDFs obtained via classical numerical diagonalization and VQE experiments on the IBMQ Manila machine for FPE matrices of different dimension using ZNE and TREX for $a = \pm 1$ and truncations $N = 2$ and $4$. Regions where the distribution is negative are indicated with dotted purple lines. Black and blue lines in (a) lie underneath red.  Black and blue overlap in (b).}
  \label{ZNE TREX comparison}
\end{figure}

PDFs obtained with VQE + TREX are shown in Fig. \ref{ZNE TREX comparison}. 
Although TREX performs at par with ZNE for the $N = 2$ ($a = 1$) case, slight deviations between the TREX and ZNE results are observed near the region of bistability for the $N = 2$ ($a = -1$) case. It is furthermore observed that for the $N = 4$ ($a = 1$) case, TREX almost performs at par with ZNE, although the performance of VQE with TREX for the $N = 4$ ($a = -1$) case is sub-optimal, to the extent that VQE with TREX fails to reproduce the region of bistability. This may be due to the relatively deep variational ans\"atz (4 repetition layers).  

\section{Discussion}
\label{sec:discussion} 

The statistics of classical dynamical systems can, in principle, be obtained from current-NISQ computers. To obtain the steady-state probability distribution, we dimensionally reduce the linear FPE differential operator to a matrix form amenable to quantum computation.  The viability of using QPE and VQE for computing the zero-mode of the linear FPE matrix on near-term quantum hardware is investigated. An accurate approximation to the steady-state PDF is obtained with the sign-corrected QPE zero-mode recovered on the local simulator. VQE on IBM Manila yields fairly accurate results the PDF when error mitigation protocols are employed to combat the effects of noise. The accuracy of VQE fluctuates with the number of iterations. TREX does not mitigate errors as effectively as ZNE for the problem considered here.  

Other studies dealing with deterministic mechanical systems on near-term quantum hardware have appeared in recent years \cite{Giannakis2019, Freeman2023, Slawinska2022, Joseph2020}.  These alternative approaches simulate the dynamics, whereas we describe a method to directly access the steady-state statistics. The application of either framework depends on the system and goal in question. That is, the relevant methods to be applied depend on whether one is interested in computing a system's dynamics or its corresponding statistics. In particular the Koopman -- von Neumann approach has been integrated with dynamical mode decomposition (DMD), an algorithm that enables the dimensional reduction of large data sets \cite{Schmid2010}, thus allowing for efficient computation. For dynamical systems with a pure point spectra (a spectrum consisting of discrete eigenvalues), the Koopman evolution operators associated with such systems can be approximately modeled and/or simulated on near-term quantum hardware using low-depth quantum circuits, thus potentially offering an exponential computational advantage over existing classical algorithms \cite{Freeman2023}. This is most readily seen when the associated Koopman-von Neumann matrix is sparse \cite{Joseph2020}. Efficient compilation of quantum programs using a low circuit depth on near-term quantum hardware also plays a key role in obtaining improved results, as demonstrated in Ref. \cite{Shi2021}. 

\section{Conclusions and Outlook}
\label{sec:conclusions} 

A unique feature of our approach is the use of direct statistical simulation (DSS), which eliminates the need to integrate differential equations over time, and provides new avenues for modeling nonlinear dynamical systems on quantum hardware.  We demonstrate that the probability distribution function of a prototypical classical nonlinear dynamical system can be recovered on near-term quantum hardware. Low-order moments of the steady-state probability distribution function such as $\langle x^2 \rangle$ can thus also be found using near-term quantum hardware. Potential future applications of the approach introduced in this paper include climate modeling \cite{Palmer2023} and the study of thermal (turbulent) convection flows at moderate and high Rayleigh numbers \cite{Pfeffer2022paper2, Pfeffer2023paper1}. Successful extension to these higher dimensional problems will certainly require improvements in variational algorithms and hardware. 

A natural question to ask is whether or not a quantum advantage over existing classical algorithms such as exact classical diagonalization can be obtained via the approaches implemented in this work. We hope to report a detailed analysis of the scaling of classical versus quantum implementations of solutions to nonlinear partial differential equations in a future work. Given the difficulties inherent in scaling up quantum computation \cite{Waintal2024}, it is possible that the most important outcome of this work is to suggest new \emph{classical} variational algorithms to find the stationary statistics of the Fokker-Planck equation.  By mapping the problem to a quantum one, methods borrowed from quantum chemistry or quantum matter physics may now be brought to bear in the context of classical dynamical systems.

\section*{Acknowledgements}
We thank Alan Bidart, Tilas Kabengele, and Daniel Lidar for useful discussions. We used IonQ machines for preliminary work, and the IBMQ Experience for the results presented here. The views expressed here are those of the authors and do not reflect the official policy or position of IBM or the IBMQ team. J. B. M. and B. M. R. acknowledge seed funding from the Brown University Office of the Vice-President for Research for this project. 

\section*{Code availability}
The following GitHub repository contains code to implement the classical (construction of the Hermitian form of the $\hat{L}_{\text{FPE}}$ matrix) and quantum (QPE + VQE) subroutines: \href{https://github.com/YashLokare02/ClassicalNonlinearSystems/tree/main}{ClassicalNonlinearSystems}.  


\providecommand{\noopsort}[1]{}\providecommand{\singleletter}[1]{#1}%

\end{document}